\documentclass{mem}

\usepackage{natbib}
\usepackage{txfonts}
\usepackage{balance}
\usepackage{graphicx}
\usepackage[a4paper]{hyperref}
\idline{00}{89}

\def\Brg{{Br$\gamma$}}
\def\df{{$\Delta\theta$}}
\def\mag{mag}
\def\deg{{$^\circ$}}
\def\MK{{M$_\mathrm{K}$}}
\def\asec{{$^{\prime\prime}$}}

\begin{document}

\title{Estimation of pattern speed in external galaxies using very young,
  stellar clusters\thanks{Based on observations collected at the European
    Southern Observatory, Chile (ESO programme 278.B-5042)}}

\author{P.~Grosb{\o}l\inst{1} \and H.~Dottori\inst{2}}

\offprints{Preben~Grosb{\o}l; \email{pgrosbol@eso.org}}

\institute{European Southern Observatory,
  Garching bei M\"unchen, Germany
\and
  Instituto de F\'{i}sica, Universidad Federal do Rio Grande do Sul,
  Porto Alegre, RS, Brazil
}

\authorrunning{Grosb{\o}l and Dottori} 

\titlerunning{Estimation of pattern
  speed using very young, stellar clusters}

\abstract{In the 60's, Prof. Str\"{o}mgren proposed to use space
  velocities and ages of moderately young stars to compute their
  places of formation in order to study the spiral structure in the
  Galaxy.  We have extended this idea to very young stellar clusters
  in nearby disk galaxies.  Near-infrared (NIR) $K$-band images of
  grand-design spiral galaxies often show bright knots along their
  spiral arms. Such knots in NGC~2997 have been identified as massive
  stellar clusters with ages of less than 10~Myr using $JHK$
  photometry and $K$-band spectra.  Ages of these clusters can be
  estimated from $JHK$ photometry.  Their azimuthal distances from the
  spiral arms, as measured in the $K$-band, correlates with their ages
  suggesting that the pattern speed of an underlying density wave can
  be derived.  This method is tested on the grand-design spiral NGC
  2997 using VLT data.
}

\maketitle{}

\section{Introduction}

\citet{stromgren63} suggested to use the migration of young stars with known
ages and space velocities to estimate the pattern speed of a density
wave in our Galaxy \citep{lin64, yuan69}.  In external galaxies where
individual stars cannot be observed, one may consider to use
integrated properties of blue, young objects (such as HII regions and
OB associations) which often are concentrated in the arms of
grand-design spiral galaxies.  Strong and very varying attenuation by
dust in the arm regions make it difficult to determine reliable,
intrinsic colors of such very young objects in visual bands. In the
NIR, several spiral galaxies had bright knots along their spiral arms
\citep{gp98}.  Such knots in NGC~2997 were identified as very young stellar
cluster (ages $<$10~Myr) using $K$-band spectra obtained with ISAAC/VLT
\citep{Grosbol06}.

It is possible to estimate accurate ages for clusters younger than
$<$8~Myr from their integrated NIR colors and/or \Brg\ emission due to
the rapid evolution of their high mass stars and the low attenuation
by dust.  Although space velocities for the clusters cannot be
obtained, the general rotation curve of the host galaxy provide enough
information to make a crude estimate of their birthplaces assuming
that they follow roughly circular orbits.

A further advantage of using NIR observations is that phase and shape
of a density wave can be measured directly.  This opens the way for
estimating its pattern speed by comparing ages of individual clusters
with their azimuthal offset relative to the spiral perturbation.  In
addition, one may study star formation induced by such density
variations e.g., through large-scale shocks or compressions in the gas
\citep{roberts69}.  In the current paper, we study the spatial and
color distributions of young stellar clusters in the southern arm of
NGC~2997 in order to test the feasibility of this scheme to estimate
the pattern speed in external galaxies.

\begin{figure}[t!]
\resizebox{\hsize}{!}{\includegraphics[clip=true]{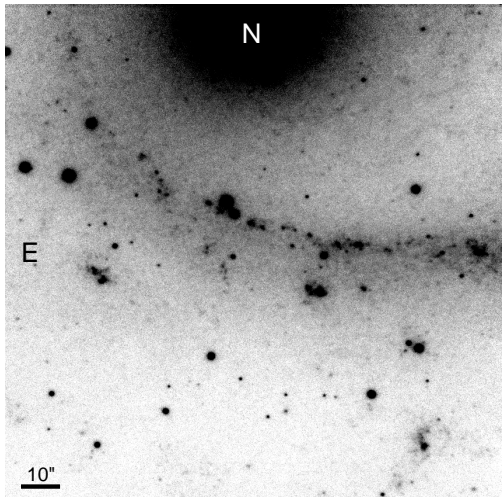}}
\caption{\footnotesize 
  Direct $K$-band image of the field around the southern arm of
  NGC~2997 used for the analysis. The scale is indicated by the bar in
  the lower left corner. }
\label{n2997k}
\end{figure}

\section{Data and reductions}

Deep $JHK$-\Brg\ images of a field around the southern arm of NGC~2997
(see Fig.~\ref{n2997k}) were obtained with ISAAC/VLT in order to study
the properties of stellar clusters.  The $K$-band map reached a
limiting magnitude of $K = 19$\,\mag\ (with errors of $\pm$0.1\,\mag)
corresponding to \MK\ = --11.3\,\mag\ (assuming a distance of
11.6~Mpc) while the linear resolution was 45~pc with a seeing of
0\farcs8 measured on the final stacked images.  Besides the
broad-band colors, a \Brg\ index (\Brg--$K$) was estimated with a zero
point defined by the average value for foreground stars in the field.

\begin{figure}[t!]
\resizebox{\hsize}{!}{\includegraphics[clip=true]{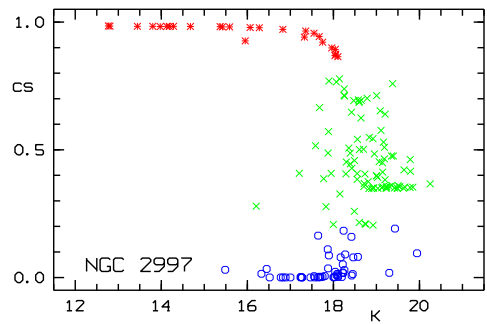}}
\caption{\footnotesize 
  The class\_star classifier, cs, as a function of apparent $K$-band
  magnitude for all sources detected by SExtractor in the NGC~2997
  field. Different source groups are indicated by symbols: diffuse
  (open circles), compact (crosses), and star-like (asterisks). }
\label{n2997cs}
\end{figure}

Sources were identified on the $K$-band image using SExtractor
\citep{bertin96} after which aperture magnitudes (diameter of 2\asec)
were measured at these source locations for all filters.  The
SExtractor class\_star classifier (cs), which range from 0 for diffuse
sources to 1 for point-like targets, is shown in Fig.~\ref{n2997cs}.
Sources were divided into 3 groups: diffuse (cs$<$0.3), compact
(0.3$<$cs$<$0.8), and star-like (0.8$<$cs).  The diffuse sources are
well separated from the compact ones and on average brighter.

\section{Spatial and color distributions}

\begin{figure}[t!]
\resizebox{\hsize}{!}{\includegraphics[clip=true]{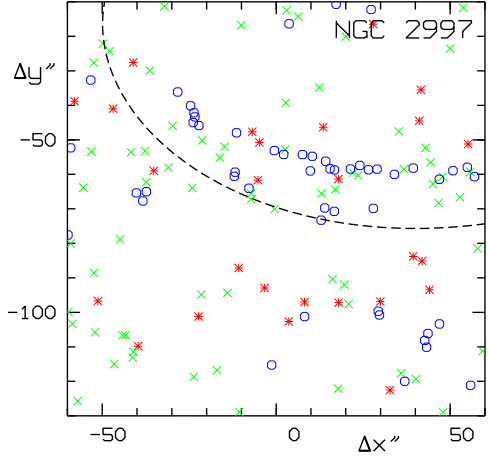}}
\caption{\footnotesize 
  All objects detected on the $K$-band image of the NGC~2997 field.  The
  dashed line indicates the average location of the two-armed spiral
  pattern. For symbols see Fig.~\ref{n2997cs}. }
\label{n2997sex}
\end{figure}

The spatial locations of the sources are shown in Fig.~\ref{n2997sex}
for the NGC~2997 field (see Fig.~\ref{n2997k}).  The average position
of the southern spiral arm (with a pitch angle of --22\fdg6) is
indicated by a dashed line which outline the phase of the $m=2$ Fourier
component of the azimuthal $K$-band intensity variation as a function of
radius.  It is clear that only the diffuse sources (open circles)
concentrate along the spiral on its inner side while the other types
of objects have a more uniform distribution.

\begin{figure}[t!]
\resizebox{\hsize}{!}{\includegraphics[clip=true]{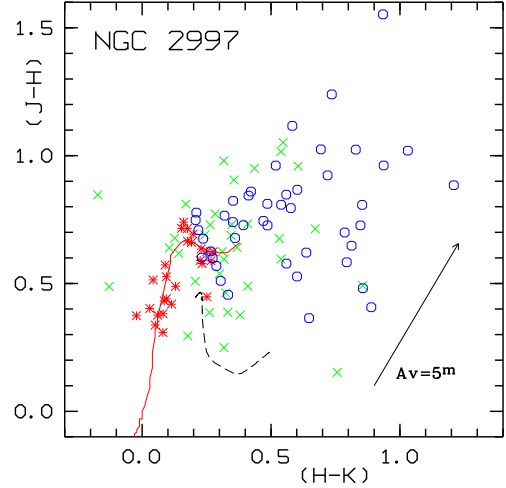}}
\caption{\footnotesize 
  Color-color diagram for all sources with color errors $<$0.1\,\mag\ in
  the NGC~2997 field. The full drawn line indicates the stellar main
  sequence while the dashed line shows a standard Starburst99
  evolutionary track for stellar clusters with continuous star
  formation for ages up to 50~Myr.  The galactic reddening vector is
  indicated for $A_V=5$\,\mag. Symbols are the same as in
  Fig.~\ref{n2997cs}. }
\label{ccd}
\end{figure}

The $(H-K)-(J-H)$ diagram for the sources is given in Fig.~\ref{ccd}
where the arrow indicates the standard galactic reddening vector.
Galactic foreground stars are located along the stellar main sequence
as outlined by the full drawn line.  An evolutionary track for stellar
clusters with continuous star formation was computed using Starburst99
\citep{leitherer99} and plotted with a dashed curve. The model had a
Salpeter IMF, an upper mass of 120~$M_\odot$, and solar metallicity.
The track starts around $(H-K) = 0.5$\,\mag\ after which $(H-K)$
becomes bluer until around 7~Myr when $(J-H)$ gets redder.  The colors
do not change much after 15-20~Myr when they approach those of
globular clusters at (0.2, 0.7). Models with instantaneous star
formation are somewhat bluer and reach slightly higher $(J-H)$ values.
The non-stellar sources are scattered above the track suggesting
typical absorptions by dust of the order of $A_V = 5$\,\mag\ but with
a large spread. The complex geometry of stars and dust in these
regions make it possible that the inner parts of the clusters have
much higher attenuation than indicated by the average value. The red
diffuse sources have colors compatible with ages $<$10~Myr.  The
compact sources are on average 2\,\mag\ fainter and have larger errors
in their color indices.

A reddening {\sl free} color index $Q = (H-K) - 0.59 \times (J-H)$ may be
computed assuming a standard galactic reddening law and a screen
model.  Starburst99 models suggest that zero-age clusters have $Q$
values around 0.4\,\mag\ and then decreases to --0.1\,\mag\ at 10~Myr
after which the index remains almost constant.  A few sources have
lower $Q$ values which may suggest that either the reddening law
differs slightly or the screen model is inappropriate due to the
complex star-dust geometry \citep{witt92, pierini04}.  There is an
anti-correlation between $Q$ and \Brg\ indices for diffuse sources as
expected from Starburst99 models.  Due to the smaller $S/N$ ratio for
\Brg, the $Q$ index was preferred as an age indicator.

\section{Locations relative to spiral arms}

\begin{figure}[t!]
\resizebox{\hsize}{!}{\includegraphics[clip=true]{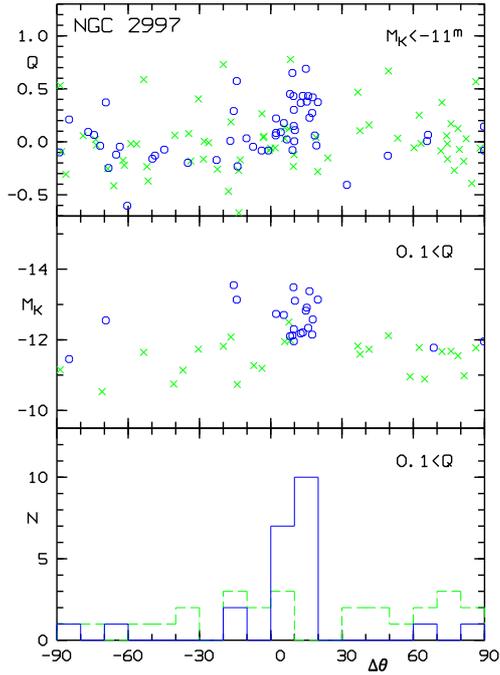}}
\caption{\footnotesize 
  Distribution of non-stellar sources as a function of their azimuthal
  distance \df\ from the two-armed spiral intensity maximum. The top
  panel shows $Q$ for sources with \MK$<$--11\,\mag, the middle panel
  displays \MK\ for young sources with $Q>0.1$\,\mag, while the button
  panel gives the histogram of young sources. The same symbols as in
  Fig.~\ref{n2997cs} are used. }
\label{dfQ}
\end{figure}

The distribution of diffuse and compact sources relative to the spiral
arms is displayed in Fig.~\ref{dfQ} using the azimuthal distance
$\Delta\theta$ from the mean two-armed spiral as derived from a
Fourier analysis of a $K$-band map \citep{Grosbol04}.  The diffuse
sources are strongly concentrated in a region 0\deg\--20\deg\ inside
the intensity maximum of the spiral arm while the compact objects have
a more uniform distribution. The sharpness of the peak of young,
diffuse objects (open circles) and its offset with respect to the
spiral arm intensity maximum (i.e., \df=0\deg) indicate a strong star
formation activity just inside the spiral arms possibly associated to
non-linear compression of gas in a spiral potential \citep{roberts69,
  gittins04}.

The concentration in the arm region is also seen clearly in the plot
of absolute magnitudes \MK\ for young clusters with 0.1\,\mag\,$<Q$ as a
function of their azimuthal offset \df\ where a sharp transition
occurs around \MK\ = --12\,\mag.  The transition could be associated
to the expulsion of gas and dust from the clusters when the first
supernovae explode \citep{bastian06, goodwin06}.  This would make the
clusters appear more compact, disrupt star formation and therefore
lead to a more rapid fainting of them.  It is noted that only young
diffuse objects (i.e., 0.1\,\mag\,$<Q$) are located in a narrow region
just inside the spiral arm whereas the compact sources have a uniform
distribution.  This suggests that formation of the most massive
clusters requires a triggering by a compression of gas associated to
the spiral arm while less massive clusters are formed more randomly
over the galactic disk.

The top diagram of Fig.~\ref{dfQ} displays the relation between
azimuthal offset and $Q$ index (age for 0.1\,\mag\,$<Q$) for the
non-stellar sources.  Compact sources (crosses) and those with
$Q<0.1$\,\mag, for which $Q$ cannot be used as an age indicator, are
distributed uniformly. On the other hand, the distribution of very
young diffuse sources (open circles) suggests a correlation where the
youngest objects are further away (inside) from the spiral arm defined
by the $K$-band intensity variation.  A rough estimate of the gradient
indicates that a 10\deg\ change in \df\ corresponds to a variation in
Q of 0.2--0.3\,\mag\ with a visual best fit of 0.22\,\mag/10\deg.  The
scatter may be caused by several factors such as photometric errors,
and formation of clusters in a region with finite azimuthal width.
Further, clusters over a significant radial range have been used (to
improve statistics) which smears out the relation due to the slightly
different pitch angles of the spiral potential and the associated star
forming front as outlined by the dust lanes \citep{grosbol99}.

\section{Birthplaces and pattern speed}

Birthplaces of the youngest clusters can be estimated using the
rotation curve which is almost flat at 185 km\,s$^{-1}$ at the radii
of the clusters \citep{peterson78}. This corresponds to a motion of
190~pc or 3\farcs9 per Myr.  Assuming that all clusters are formed in
a star forming front with the same shape as the density wave (just
offset from it), a first order estimate of the pattern speed can be
given as: $$\Omega_{\rm p} = (V_c - V_r/\tan(i))/r -
\delta\theta/\delta a $$ where $V_c$ is the circular velocity, $V_r$
the radial velocity, $r$ the average radius of the clusters, $i$ the
pitch angle of the spiral (negative for trailing patterns), and
$\delta\theta$ the change in azimuthal offset
\df\ for clusters with an age difference of $\delta a$.  This
approximation is valid only for $\delta a \ll \pi/2\kappa \approx
50$~Myr for NGC~2997 where $\kappa$ is the epicyclic frequency.

The estimate of $\delta\theta/\delta a$ can be derived from the top
diagram in Fig.~\ref{dfQ} which suggests $\delta\theta/\delta Q =
0.22$\,\mag/10\deg.  The Starburst99 model indicates a $\delta
a/\delta Q$ of --22~Myr per mag for the range of $Q$ around
0.2\,\mag\ with a slightly more shallow slope for models with upper
mass limits below 30~M$_\odot$.  This yields $\delta\theta/\delta a
\approx 36$ km\,s$^{-1}$\,kpc$^{-1}$ and $\Omega_{\rm p} = (54 - 36)$
km\,s$^{-1}$\,kpc$^{-1}$ = 18 km\,s$^{-1}$\,kpc$^{-1}$ for pure
circular velocities while $V_r = -10$ km\,s$^{-1}$ would lower the
estimate to around 12 km\,s$^{-1}$\,kpc$^{-1}$. Both estimates of the
pattern speed place the inner Lindblad resonance (ILR) close to the
inner radius of the main spiral pattern and co-rotation significantly
outside the symmetric pattern in good agreement with other
determinations.

\section{Conclusions}

The analysis of sources in a field centered on the southern arm of
NGC~2997 observed in NIR bands with ISAAC/VLT offers the following
tentative conclusions:

\begin{itemize}
\itemsep=0pt

\item the youngest stellar clusters with magnitudes \MK\ $<-12$\,\mag\ are
  concentrated in the arm regions,

\item the bright, young clusters (0.1\,\mag\,$<Q$) show an age gradient where the
  youngest are further away (in front of) the spiral arm,

\item the pattern speed of the density wave can be estimated to $\approx$18
  km\,s$^{-1}$\,kpc$^{-1}$ using this gradient and assuming circular
  motions and formation along a front with the same shape as the
  spiral arms, and

\item fainter young clusters have a more uniform distribution.

\end{itemize}

This is consistent with the standard density wave picture where the
main spiral pattern starts just outside the ILR.  The most massive
clusters seem to be triggered by a front associated to the density
wave while smaller clusters are more uniform distributed.  The use of
broad band NIR photometry to estimate ages of very young stellar
clusters in spiral galaxies and from those calculate birthplaces
provides a new independent method to estimate the pattern speed of
density waves in nearby, grand-design spirals.

%%% BIBLIOGRAPHY

\bibliographystyle{aa}

\end{document}